\documentclass[acmsmall,authorversion,nonacm]{acmart}   

\AtBeginDocument{%
  \providecommand\BibTeX{{%
    \normalfont B\kern-0.5em{\scshape i\kern-0.25em b}\kern-0.8em\TeX}}}

\usepackage{url}
\usepackage{doi}

\usepackage{currfile}
\usepackage{textcomp}
\usepackage{enumitem}

\begin{document}

\title{A Contract Service Provider Model for Virtual Assets}

\author{Thomas Hardjono}
\affiliation{%
\institution{MIT Connection Science}
\city{Cambridge, MA 02139}
}
\email{hardjono@mit.edu}

\author{Alexander Lipton}
\affiliation{%
\institution{MIT Connection Science}
\city{Cambridge, MA 02139}
}
\email{alexlip@mit.edu}

\author{Alex Pentland}
\affiliation{%
\institution{MIT Connection Science}
\city{Cambridge, MA 02139}
}
\email{pentland@mit.edu}



\begin{abstract}
With the recent rise in the cost of transactions on blockchain platforms
there is a need to explore other service models
that may provide a more predictable cost to customers and end-users.
We discuss the {\em contract service provider} (CSP) model
as a counterpart of the successful Internet ISP model.
Similar to the ISP business model based on peered routing-networks,
the CSP business model is based multiple
CSP entities forming a {\em CSP community} or group
offering a contract service for specific types of virtual assets.
We discuss the {\em contract domain} construct
which encapsulates well-defined smart contract primitives, policies and contract-ledger.
We offer a number of design principles borrowed
from the design principles of the Internet architecture.
\end{abstract}





\maketitle




\section{Introduction}
\label{sec:Introduction}

Blockchains and distributed ledger technology (DLT) currently face a number of growing pains,
one being the challenge of providing a suitable cost and pricing model
for entities that participate in a blockchain ecosystem.
The recent rise in the cost of transactions in 
Bitcoin and Ethereum
illustrates the reality that the virtual asset industry 
is still in its nascency -- even with Bitcoin being over a decade old.
This rise in costs is exemplified not only by Bitcoin,
but also by Ethereum with the recent hype in ``decentralized finance'' 
(DeFi)~\cite{VoellFoxley2020-coindesk,CawreyKeoun2020-coindesk}.
This points to the need to consider alternative
service pricing models that offer stability over time
and affordability to ordinary end-users.

The current work proposes a new kind of service provider 
for blockchain-based asset related services,
which we refer to as the {\em Contract Service Provider} (CSP).
This contract-centric view of services follows from
the concept of Function-as-a-Service (FaaS)~\cite{AlderAsokan2018-Secure-FaaS,Fowler-Serverless-FaaS},
and it represents a further logical narrowing
of the notions of software-as-a-service (SaaS) and the variations
of the hosted computing model (e.g. container-as-a-service, node-as-a-service, etc.).

The current work provides discussions around three (3) areas or themes of focus,
namely that:
\begin{enumerate}

\item	Blockchain interoperability architectures that support the
mobility of virtual assets across
blockchain networks tempers or reduces fluctuations in transaction costs or fees.

\item	Decentralization infrastructures (i.e. blockchain nodes) must 
operate agnostically to the economic value of the virtual assets
flowing through the nodes.
This is one of the important corollaries of the end-to-end principle~\cite{SaltzerReed84,Clark88}
of the Internet architecture -- which when implemented together with
interoperable blockchain architectures permits true scaling of blockchain-based services.

\item	Simple and well-defined smart contracts permits the standardization of contract primitives
for specific types of assets.
This, in turn, permits smart contracts to be accessible
as a Function-as-a-Service made available by contract service providers.

\end{enumerate}

The overall goal of the CSP model is to simplify end-user access to services dealing
with virtual assets,
and supporting fixed fees to end-users for contract invocations (e.g. subscription model based on tiered pricing).
The contract primitives are ``light weight'' in that
it provides only a limited set of operations (in the sense of Bitcoin's limited set of op-codes),
each of which consumes a deterministic number of CPU operations.

We define CSPs more precisely in Section~\ref{sec:CSP-Communities},
and discuss the notion of the contract-domain in Section~\ref{sec:ContractDomain}.
We discuss the use of well-defined contract primitives in Section~\ref{sec:ContractPrimitives}
as the basis of services.
In order to adhere to the end-to-end principle,
we discuss the importance of externalizing the value of the virtual assets to the blockchain infrastructure
in Section~\ref{subsec:SourcesVirtualAssets}.
We close the paper with some conclusions in Section~\ref{sec:Conclusions}.

We seek to make the current paper readable to a broad audience,
and as such have sought to limit the usage of technical jargon.
However,
we assume the reader is at least familiar with Bitcoin, cryptocurrencies,
and distributed ledgers generally.

\section{Interoperability, Asset Mobility and Transaction Costs}
\label{sec:interop}

We believe the rise in transaction costs would be tempered if blockchain systems had the same degree
of interoperability as IP routing autonomous systems.
If virtual assets had ease of mobility or transferability across
different blockchain systems, 
then end-users and asset holders
would have the freedom to move their assets elsewhere should
their current blockchain system become too costly to do business in.
In general,
blockchain interoperability is crucial not only for asset mobility,
but also for the scalability and survivability of blockchain networks 
as a whole~\cite{HardjonoLipton-IEEETEMS-2019}.

\begin{figure}[t]
\centering
\includegraphics[width=1.0\textwidth, trim={0.0cm 0.0cm 0.0cm 0.0cm}, clip]{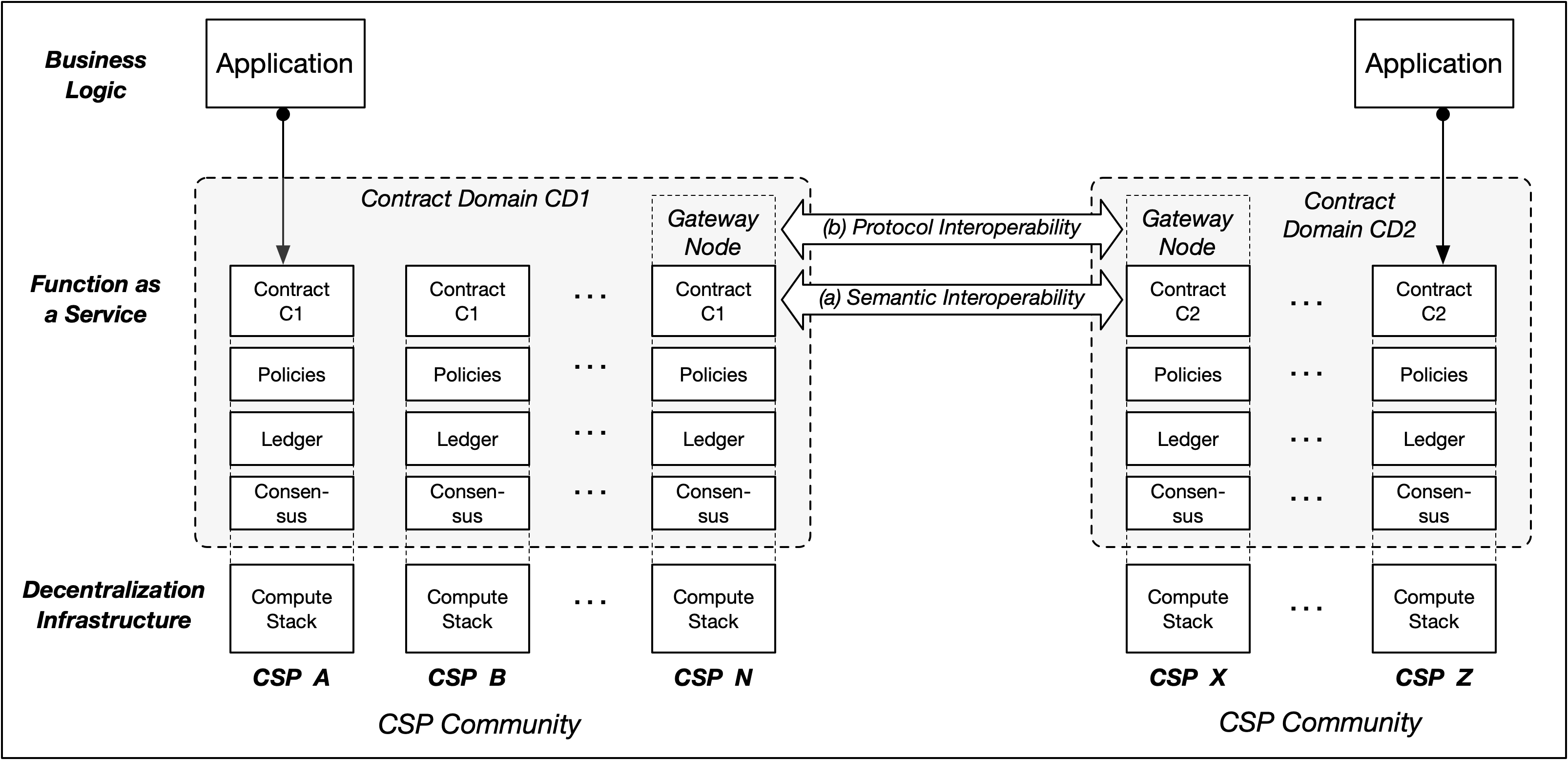}
\caption{Interoperability of contract semantics and gateway protocols}
\label{fig:SemanticInterop}
\end{figure}

\subsection{Interoperability at Peer Layers}

In order for asset mobility to occur seamlessly across blockchain systems,
interoperability needs to occur at several levels 
in the blockchain functional layers (see Figure~\ref{fig:SemanticInterop}):

\begin{itemize}

\item	{\em Interoperability of contract semantics}:
When virtual assets are to be transferred from one blockchain system
to another, both blockchain systems must
have the same semantic understanding of the meaning of the ``transfer'' transaction.
This must be true regardless of the specific syntax of the contract language
being used in either of the blockchain systems~\cite{HazardHardjono2016}.
This is illustrated as item (a) in Figure~\ref{fig:SemanticInterop}.

For example,
when an Originator (Alice) seeks to transfer her virtual asset
from its current blockchain system (e.g. B1) 
to an a Beneficiary (Bob) in a different blockchain system (e.g. B2),
the smart contract implementing this unidirectional
cross-chain transfer must have compatible (or identical) state-machines -- which must
result in the states maintained by their respective ledgers to achieve
consistency across systems 
(independent of whether the ledgers are permissioned or permissionless).
Semantically, the asset in blockchain B1 must be extinguished
and then introduced into in blockchain B2
in an atomic manner while preventing double-spending by Alice
while the transfer is being carried-out.

\item	{\em Gateway protocol interoperability}:
The nodes that perform asset mobility related functions -- referred to as {\em gateway nodes} --
must implement the same gateway protocols in order for both ledgers
to arrive at the same consistent state~\cite{IETF-draft-hardjono-gateways}.

Using the previous Alice and Bob example of the cross-chain asset transfer,
the gateway G1 in the origin blockchain B1
must execute the same technical transfer protocol as gateway G2 in the destination blockchain B2.
This is illustrated as flow (b) in Figure~\ref{fig:SemanticInterop}.
Since it is reasonable to assume that different 
technical transfer protocols may be developed and standardized in the future,
gateways G1 and G2 must first negotiate the common protocol (type and version)
supported by both gateways.
Protocol-type and parameter negotiation is a common phase
in many Internet protocols (e.g. TLS handshake negotiation).

\item	{\em Common reference to authoritative asset profile}:
Virtual asset mobility requires an authoritative definition of the virtual asset in question.
We refer to this metadata as the {\em asset profile}.
It is essentially a prospectus file of a regulated asset that includes
information and resources describing the virtual asset.  
The asset profile is independent from the specific
instantiation of the asset (on a blockchain or otherwise) and
independent from its instance-ownership information~\cite{IETF-draft-hardjono-gateways}.

When the Originator (Alice) in blockchain B1
seeks to transfer virtual assets to a Beneficiary (Bob) in blockchain B2,
both Alice and Bob must have the means to refer to 
the same asset definition (namely the asset profile file).
Consequently, the gateway nodes G1 and G2 that are performing the transfer on behalf of Alice and Bob
must also have the means to refer to (point to, or hash of) the asset profile file.
A copy of the signed asset profile file may be represented on-chain,
or it may be elsewhere off-chain
(e.g. at a well known asset depository~\cite{DTCC-Project-Whitney-2020}).
We discuss this further in Section~\ref{subsec:SourcesVirtualAssets}.

\end{itemize}

\subsection{Entities in the Ecosystem}

Figure~\ref{fig:SemanticInterop} incorporates a number of entities
and functions that we will discuss further in the ensuing sections.
We provide a brief definitions of these here
in order to facilitate discussion.
\begin{itemize}

\item	{\em Contract}: The Smart contract service
defined by the set of primitive operations implemented on-chain and the asset-types
processed through the primitives.
Figure~\ref{fig:SemanticInterop} shows two contracts C1 and C2,
which are implemented by the nodes participating in the Contract Domain CD1 and CD2 respectively.

\item	{\em Contract Service Provider} (CSP): 
A regulated service provider (e.g. registered business)
that participates in making available contract services to its customers.
Note that a CSP is in fact also Virtual Asset Service Provider (VASP)
as defined under the FATF definition and the Travel Rule 
(see~\cite{FATF-Recommendation15-2018,FATF-Guidance-2019}).

\item	{\em CSP Community}: A group of CSPs
offering contract services for a given asset type
on a blockchain network consisting of their nodes.

\item	{\em Contract Domain}: The various computing resources
required to implement a contract service,
including contract-specific functional components 
(i.e. primitives, contract-ledger, consensus algorithm and domain policies) 
and the other technological constructs that implement the domain.

\item	{\em Gateway node}: The node belonging to a CSP that performs
cross-domain (cross-chain) asset transfers.

\end{itemize}

An overall summary of Figure~\ref{fig:SemanticInterop} is as follows.
A group of CSPs (labelled $A, B,\ldots, N$) form a CSP community
that agrees to make available contract $C1$ on their nodes.
The decentralization functions (contract, policies, ledger and consensus protocol)
used to implement the contract form a contract domain CD1.
Each member of the CSP community is free to use different
compute stack technologies (e.g. bare metal, hosted node, container as a service, etc.)
to stand-up their nodes that participates in the contract domain.
Another CSP community is also shown in Figure~\ref{fig:SemanticInterop},
consisting of CSPs labelled $X,\ldots, Z$,
which agree to make available contract $C2$ on their nodes forming domain CD2.
As will be discussed later,
within this contract-centric CSP model a given CSP as a business entity
may participate in several different CSP communities (contract domains) simultaneously.

\subsection{Principles from the Internet Architecture}
\label{subsec:PrinciplesInternet}

The following summarizes a number of key principles from the Internet architecture
that are relevant to blockchain interoperability:
\begin{itemize}

\item {\em Observance of the end-to-end principle}: Smart contracts and the blockchain
infrastructure must relegate the notion of economic value of assets
to the end-points of the transaction outside of the blockchain system
(i.e. the originator and beneficiary).
The principle in essence states
that any semantic meaning associated with the data (bytes) flowing through a network
must be external to the network itself~\cite{SaltzerReed84,Clark88}.
That is, it must be kept at the ``edges'' of the network at the sender and the receiver.
The network infrastructure must be agnostic as to whether a set of data packets (datagrams)
belong to one user application (e.g. email) or another application (e.g. video streaming).

When re-casted to a blockchain system consisting of multiple nodes,
the end-to-end principle demands that for cross-chain interoperability
the computing nodes of a blockchain network
be agnostic to the economic value of the virtual assets (bytes) 
being processed collectively by the nodes.
The blockchain ledger must be viewed simply as state-table with no attached economic significance,
much as a routing-table in an ISP domain has no economic significance
to the end-users.

\item {\em Observance of the autonomous systems principle}:
Contract service providers must be able to form their own communities
by collectively pooling their resources (i.e. nodes),
without being reliant on other platforms.
That is, each blockchain system must operate as a true autonomous system~\cite{Clark88}
regardless of the number of nodes in the blockchain
and the access model (permissionless or permissioned).
Standard interfaces and cross-chain protocols must be defined 
to prevent end-users' assets from suffering from ``platform capture''.
Blockchain systems must be designed so as 
not be dependent on third parties (e.g. crypto-exchange entities)
in order to achieve asset mobility.
Any such dependence reduces the autonomy of a blockchain system,
and the overall technical value of blockchain/DLT technology.

\end{itemize}

We believe the end-to-end principle (i.e. lack of adherence to it today)
contributes to the problems around fluctuating transaction costs.
A blockchain service model that operates based on an inherent tight-coupling
between the actual infrastructure operational costs 
(i.e. cost of CPU hardware, electricity, etc.)
with the tokenization of user-access
results in the unpredictability of transaction costs.
Such a model essentially makes the infrastructure (i.e. nodes) no longer agnostic
to the economic value of the ``bytes'' passing through the infrastructure.

Although this tight-coupling maybe motivated by the desire
to attain an incentives equilibrium for honest nodes processing
transactions (e.g. mining nodes in Bitcoin),
this approach appears to be unsustainable in the long term.
Node-operators (i.e. miners) and token-holders are instead motivated
to manipulate the price of tokens at the expense of other users
(e.g. indirect inflation through fake transactions~\cite{Vigna2019a,GriffinShams2019}).

\begin{figure}[t]
\centering
\includegraphics[width=1.0\textwidth, trim={0.0cm 0.0cm 0.0cm 0.0cm}, clip]{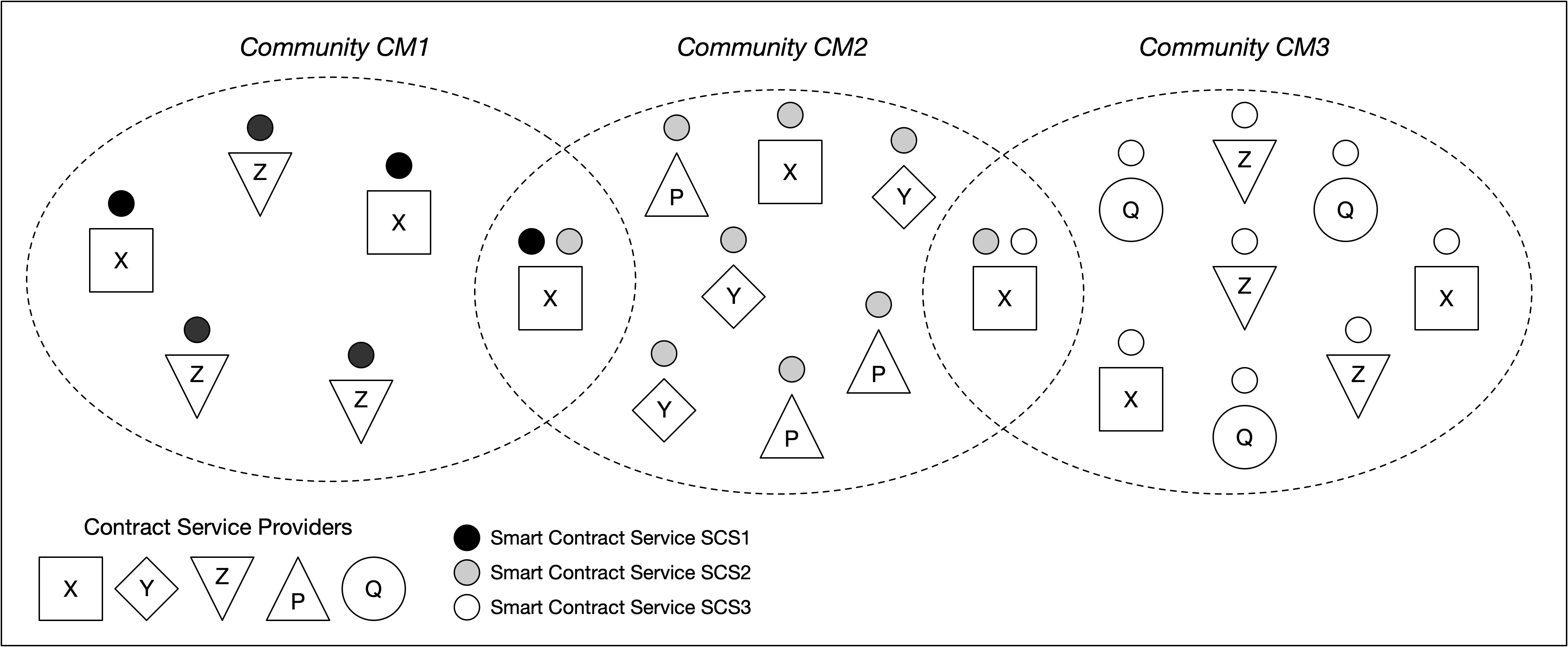}
\caption{Simple example of three (3) CSP communities viewed from a contract-centric perspective}
\label{fig:CSP-community}
\end{figure}

\section{The CSP and CSP Communities}
\label{sec:CSP-Communities}

A Contract Service Provider (CSP) is a regulated service provider
who collaborates with other CSPs
in making available on distributed nodes one or more {\em smart contract primitives} 
that consists of well-defined operations applicable to virtual assets.
Collectively,
the group of CSPs is said to offer {\em contract services} via well-defined APIs
to one or more end-users (customers),
which may include individuals and organizations.
The contract primitives and the supported types of assets are chosen by the CSPs prior to deployment.
All smart contracts employed by the CSPs are authored (programmed) by the CSPs,
and the end-users are not permitted to to load/run their on smart contracts.

As mentioned previously,
the goal of the CSP model is to simplify end-user access to services dealing
with virtual assets.
Contract invocations are based on fixed fees that are not tied to any dynamic incentives mechanisms
(i.e. proof or work).
The contract primitives must be ``light weight'' in that
it provides only a limited set of operations,
each of which consumes a deterministic number of CPU operations.

This allows a CSP to ``containerize'' node-stacks (i.e. virtual machine software images)
that implement these contract primitives
and to deploy these images in different platform configurations~\cite{HardjonoSmith2020-NodeAttest-WWJ}.
Thus, for a CSP the cost of operations are also deterministic
and node-deployment automation can be further introduced 
to reduce operational costs. 
The CSP is free to employ different resource management strategies
to implement the requirements of each contract domain
(e.g. bare-metal nodes;
virtualized nodes on its private cloud;
multi-tenanted virtualized nodes on public cloud;
SGX-secured nodes; etc.).

\subsection{The CSP Community}

We define a {\em CSP community} as a group of regulated contract service providers
offering asset-related smart contract services on a blockchain network
whose nodes are composed of the computing resources of the members of the CSP community.
The notion of a community is contract-centric
in that the CSPs in a community agree to allocate computing resources
for a given smart contract service.
A given CSP community may be constituted to provide contract services for a single smart contract
or for several related smart contracts on the same blockchain network.
A {\em contract service} is defined by the set of contract primitives implemented on-chain and the {\em asset-types}
processed through those primitives.
For a given contract domain (see below),
a customer should be associated with (belong to) only one CSP in the CSP community.
The CSP community must clearly define beforehand the contract primitives 
and the supported asset-types
that constitute the contract-service,
as well as the pricing structures
in order to provide transparency to customers.

From a legal perspective,
a CSP community must be founded on a legal contractual agreement 
that defines the obligations and liabilities of each of the member.
The community must define a service level agreement (SLA) for the customers of its membership.
This approach is akin to multi-lateral peering agreements used by Internet Service Providers (ISPs),
which defines common data routing responsibilities across multiple networks.

A simple example of three (3) CSP communities is shown in Figure~\ref{fig:CSP-community}.
Community CM1 are CSPs providing contract service SCS1,
the Community CM2 providing contract service SCS2,
while Community CM3 the contract service SCS3.
The CSP-X (shown as the square box in Figure~\ref{fig:CSP-community}) 
is active in all three communities,
where in each case it makes available computing resources (nodes)
to each community.

\subsection{CSP Community Membership Agreement}
\label{subsec:MembershipAgreement}

The computing resource to be allocated by each CSP in a CSP-community,
the specifications of the technical mechanisms (e.g. protocols) to be used in the community,
its core {\em operating rules}~\cite{VISA2017-CoreOperatingRules},
as well as other operational aspects of the contract domain
is expressed in the {\em CSP community membership agreement} document
-- which is a legally binding contractual agreement.
The community membership agreement may also specify the number of CSP entities
minimally (maximally) required to establish the contract domain,
and the methods to add or subtract the CSP membership.
The agreement may also place a time duration commitment on CSPs,
meaning that once a contract service is made operational by a CSP community
the CSPs are bound to be a member (i.e. allocate nodes and computing resources)
until the end of the duration.

From a revenue perspective,
the CSP community membership agreement must specify a fair and attractive 
revenue sharing structure for the CSP members in that community.
For example, a revenue sharing model could be based
on the number of transactions (local or cross-domain) and the size of the customer-base overall.

Although beyond the scope of the current work,
the notion of a {\em node-stack diversity index}
could be defined in the CSP community membership agreement
as a measure of the diversity of the node-implementation 
technologies~\cite{NIST-8202-2018,HardjonoSmith2020-NodeAttest-WWJ}.
The node-stack diversity index may provide customers with a tangible indicator
of the resiliency of the blockchain network as a whole
(e.g. degree of resistance against malwares/viruses
targeted to specific node softwares).

\section{The Contract Domain}
\label{sec:ContractDomain}

\begin{figure}[t]
\centering
\includegraphics[width=1.0\textwidth, trim={0.0cm 0.0cm 0.0cm 0.0cm}, clip]{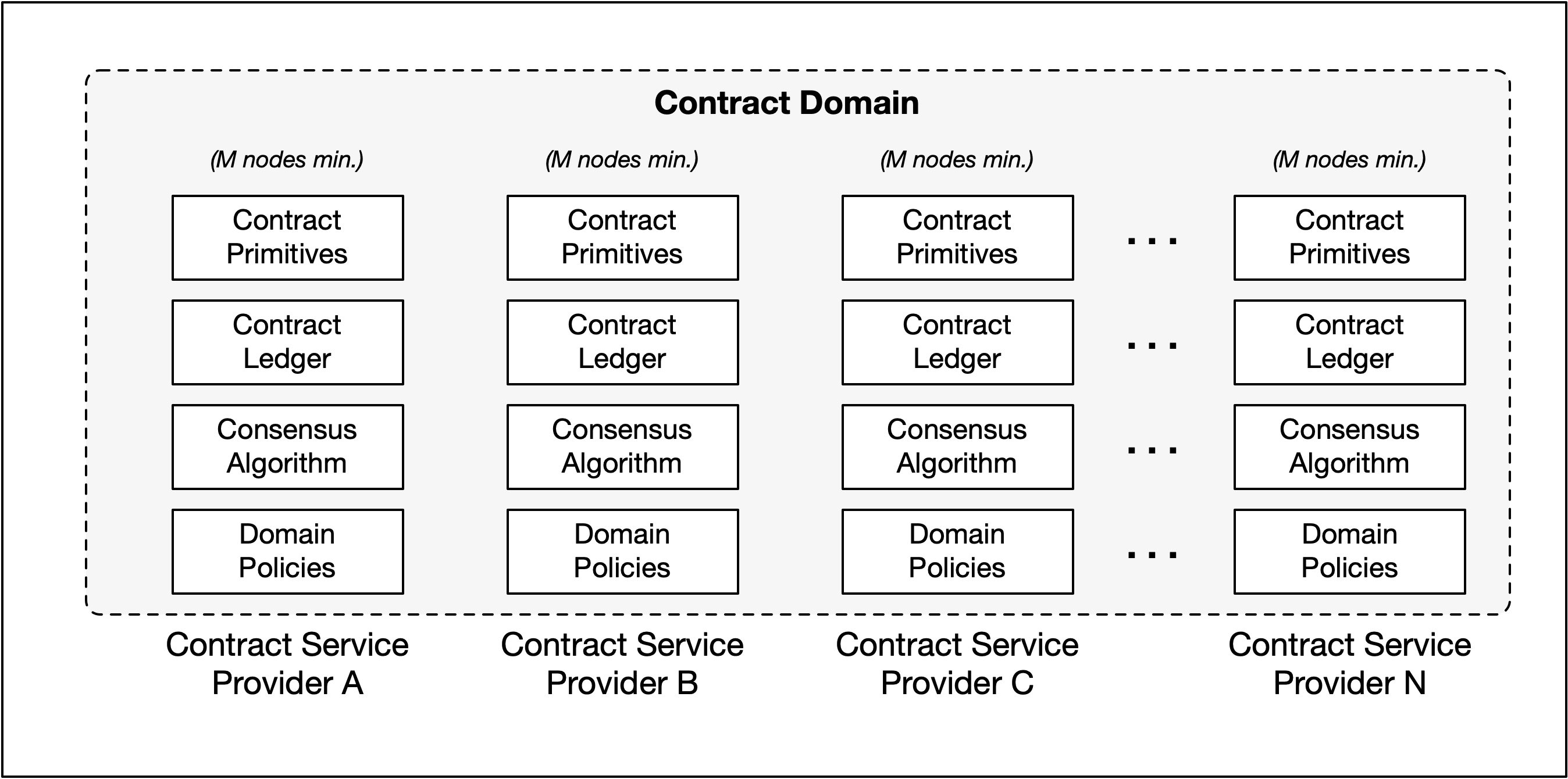}
\caption{Overview of a contract domain}
\label{fig:SimpleDomain}
\end{figure}

We use the notion of the contract domain in order to reason more accurately about the various
technical and implementation aspects of a contract service,
including access to the components of the contract service (e.g. contract calling APIs, ledger, etc).
A {\em contract domain} is defined by a CSP community through
a combination of the following:
(i) the set of contract-primitives that constitute the contract service
together with the contract ledger and consensus algorithm for the ledger;
(ii) the policies regarding the asset types permitted to be transacted in the jurisdiction of operation of the CSP community;
(iii) the nodes infrastructure that implements the contract-primitives and enforces policies.
This is illustrated in Figure~\ref{fig:SimpleDomain}.

%

A contract domain coincides with the CSP community in that both
represent the same participating business entities (namely CSPs)
and the resources (i.e nodes) dedicated by the business entities 
to establish the contract service.
Thus, for business reasons
a given CSP may simultaneously be a member of multiple CSP communities 
(each with a separate contract domain) at any given moment.
In each case, the CSP must allocate the computing resources required for each contract domain
(e.g. minimal $M$ nodes) and observe the core operating rules of each community.
In each community,
the contract domain structure ensures that a separate ledger and consensus mechanism is used to record
the asset transactions in that contract domain.

\begin{figure}[t]
\centering
\includegraphics[width=1.0\textwidth, trim={0.0cm 0.0cm 0.0cm 0.0cm}, clip]{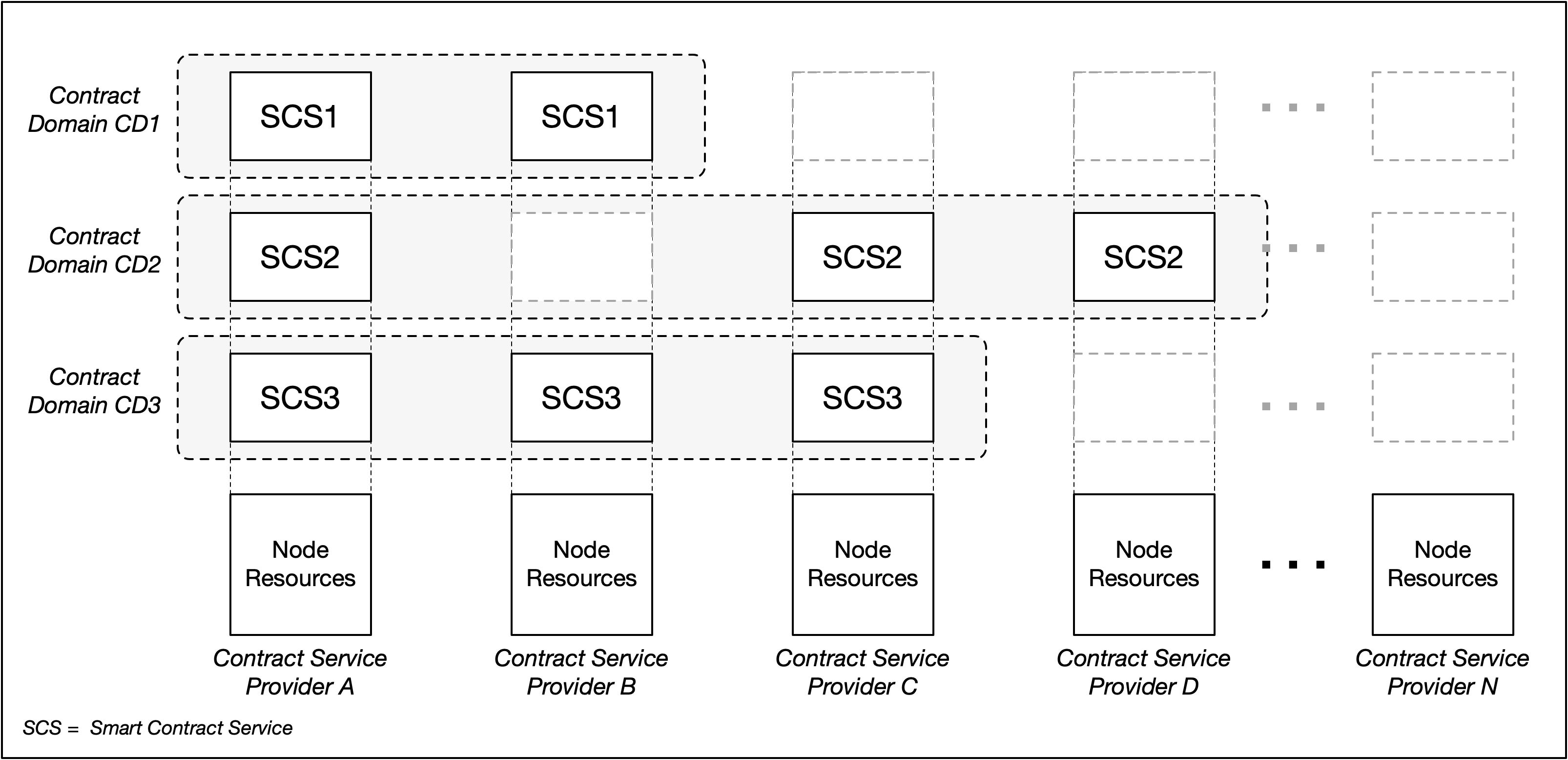}
\caption{Illustration of CSP participation in several contract domains}
\label{fig:contractdomains}
\end{figure}

A given CSP may participate in multiple CSP communities (contract domains) simultaneously,
dedicating the nodes required in each of the contract domains.
Figure~\ref{fig:contractdomains} illustrates this case.
For example,
CSP-A is shown to be participating in contract-domains CD1, CD2 and CD3,
while CSP-C is participating in domains CD2 and CD3 only.
This means that CSP-A, CSP-B and CSP-C share a common contract-ledger for CD3
and jointly participate in the consensus mechanism to maintain that contract-ledger.
The CSP-A, CSP-C and CSP-D a common contract-ledger for domain CD2.

It is important to note that when a CSP participates in multiple CSP communities
it must adhere by the membership agreement and operating rules of each community.
The membership legal agreement must prohibit {\em domain intersections}
from occurring, either inadvertently or intentionally by CSPs.
Prohibiting domain intersection means that 
a CSP must not use the same node in two (or more) contract domains,
as this may lead to the node acting as 
an unauthorized ``bridge'' between the two contract domains.

As will be discussed below,
asset transfers between two contract domains must be performed
in an explicit and transparent manner using an 
atomic cross-domain asset transfer protocol~\cite{IETF-draft-hardjono-gateways}.
The gateway nodes in the two domains
which perform the asset transfer protocol
may happen to belong to the same CSP entity, but this coincidence
(e.g. through the gateway selection algorithm in each domain)
must be transparent to all other CSP members in the two respective domains.

There are several important aspects about a contract domain
and the resources (nodes) implementing the contract domain:
\begin{itemize}

\item	{\em Per-domain consensus algorithm and per-domain ledger}:
The nodes that implement the contract service in a domain
employ a separate consensus algorithm and contract-ledger specifically for that contract service.
The choice of consensus algorithm and the structure of the ledger-blocks
are defined by the CSP community through their formative the legal agreements.
Thus, in Figure~\ref{fig:contractdomains} there is a separate ledger
for contract service SCS1, SCS2 and SCS3.
For example,
CSP-A who participates in all three domains CD1, CD2 and CD3
must put forward distinct (non-intersecting) nodes (e.g. minimal $M$ nodes per domain).
For domain CD1, the nodes establish and maintain the contract-ledger CL1,
for domain CD2 the ledger CL2
and for domain CD3 the ledger CL3.

\item	{\em Opaque ledgers and contracts}:
When an end-user customer obtains access  
to a contract service from a CSP who is a member of a CSP community,
the customer has visibility only into the relevant resources
(e.g. contract ledger) for that contract domain.
Using Figure~\ref{fig:contractdomains},
if a customer of CSP-A purchases access to contract-service SCS3 (domain CD3),
the customer has visibility only to the ledger for SCS3
(i.e. no visibility or access to ledgers for SCS1 and SCS2).

\item	{\em No domain intersections}:
A given CSP must not deploy the same node in two (or more) contract domains.
Asset transfers between two contract domains must be performed
in an explicit, transparent and non-repudiable manner.

\end{itemize}

In the remainder of this paper,
we will use the term ``CSP community'' when discussing the business and legal aspects of contract services by a CSP,
and we shall use ``contract domains'' when discussing 
the technical aspects of the contracts, nodes, ledgers and blockchain.

\section{Primitives in a Contract Service}
\label{sec:ContractPrimitives}

A core requirement of the CSP model is the well-defined, simple and modular smart contract primitives.
This ensures that efficient on-chain primitives
have a limited number of operations.
Furthermore,
this ensures that complex business logic is located off-chain within the customer application.

Although there are several possible primitives that a contract-service may use,
the following represents the basic primitives
which must be implemented by a contract domain~\cite{IETF-draft-hardjono-gateways,IETF-draft-hargreaves-ODAP}:
\begin{itemize}

\item	{\em Asset transfer from one customer to another}:
This operation moves the ownership of a virtual asset instance
from one customer to a second customer,
both of which must have been previously onboarded by a CSP member in the community.
This is equivalent to Bitcoin's payment to public key (or to a hash of public key).

\item	{\em Asset escrow to another customer or to CSP}:
This operation conditionally moves the ownership of a virtual asset instance
from one customer to another, or from one customer to a CSP in the community.
The escrows must be time-limited,
meaning that if the condition fails to be satisfied within the specified time,
the asset reverts back to the customer.

\item	{\em Asset ingress into blockchain}:
This operation introduces a regulated virtual asset instance
into the contract domain,
making it available for exchanges between customers of the CSPs in that domain.
This asset-introduction operation maybe at the request of a customer of a CSP.
This operation is available only to a CSP
because the CSP must validate the legal status of the 
asset prior to introducing it into the contract domain.
A customer cannot introduce virtual assets on their own.

\item	{\em Asset egress from blockchain}:
This operation removes (extinguishes) a regulated virtual asset instance
from the contract domain,
making it no longer available to customers of the CSPs in that domain.
This operation is available only to a CSP.
It marks the ledger to indicate that the asset has moved to another contract domain
and therefore unavailable for further use.

\end{itemize}

Other possible contract primitives include those pertaining
to key management and to different types of virtual assets.
For example,
the key management tasks include:
introduction of a customer (new customer) public-key into the contract domain;
key-rotation (or revocation) of customer's public-key;
introduction of a new CSP public-key;
key-rotation (or revocation) of CSP's public-key;
and so on.
Asset related tasks include:
introduction of a new asset-type (e.g. new stablecoin, etc.),
removal of an existing asset-type; and so on.
Cross-domain (cross-chain) primitives are further discussed in~\cite{HardjonoLipton2020-CSP-Arxiv}.

\section{Profiles of Virtual Assets and Externalization of Value}
\label{subsec:SourcesVirtualAssets}

As mentioned in Section~\ref{sec:interop},
in order for interoperability to be achieved between two contract domains
the nodes that participate in both domains respectively must have
a common definition of the virtual asset being transacted.
We refer to this authoritative definition (metadata) of a virtual asset
as its {\em asset profile}.
It is essentially a prospectus document of a regulated virtual asset that includes
information and resources describing the virtual asset.  
This includes the asset name/code, issuing authority, denomination,
date of issue, intended systems of circulation,
jurisdictions, and the URLs and mechanisms to validate the information.  
The asset profile must be digitally signed by its {\em Profile Authority} (profile issuer),
which may or may not be the same entity that creates the instantiations of the virtual asset
defined by the profile document.

From the gateway protocol interoperability perspective (see Figure~\ref{fig:SemanticInterop}),
there are a number of technical requirements to achieve interoperability across contract-domains:
\begin{itemize}

\item	{\em Externalization of value}:  The nodes and the gateway protocol 
between contract-domains must be agnostic
(oblivious) to the economic or monetary value of the virtual asset
being transferred~\cite{IETF-draft-hardjono-gateways}.

\item	{\em Separation of asset profile from value sources and technological instantiation}:
As a prospectus metadata, an asset profile document
(e.g. signed JSON file) may define the permitted forms of the asset technological instantiation
(e.g. blockchain-based, Chaumian eCash, etc).
However, the profile document must be standalone and 
be independent from its specific asset-instantiations
or evidence of instantiations.
It must also be separated from the evidence of legal ownership of the instantiations
(e.g. asset bound to an owner's public-key on a given ledger).

\end{itemize}

The asset profile metadata is used when a regulated entity 
seeks to issue digital or virtual assets
in a jurisdiction that recognizes the asset type.
The asset profile may specify the technological implementations
(e.g. blockchains, DLTs, hash-graphs, etc.)
required to instantiate the virtual asset.
The function of legally issuing virtual assets 
in a digital representation
is assumed to start outside the contract domain,
and is performed by an external entity referred to as the asset {\em Issuer} authority.
A symmetrical role is assumed to be carried out by an asset {\em Acquirer} authority,
which performs the reverse function.
The asset Issuer and Acquirer can be the same external entity.
This assumption is consistent also with a number of exploratory projects
that have been reported 
(e.g. see~\cite{MAS-Singapore-Ubin5-2020,DTCC-Project-Whitney-2020,DTCC-Project-ION-2020}).
The method to determine the value of an asset is outside of the current work,
and several mechanisms have been proposed 
(e.g. see~\cite{TascaTessone2019Ledger-Taxonomy,AnkenbrandBieri2020-CVCBT} for a proposed taxonomy).

\begin{figure}[t]
\centering
\includegraphics[width=1.0\textwidth, trim={0.0cm 0.0cm 0.0cm 0.0cm}, clip]{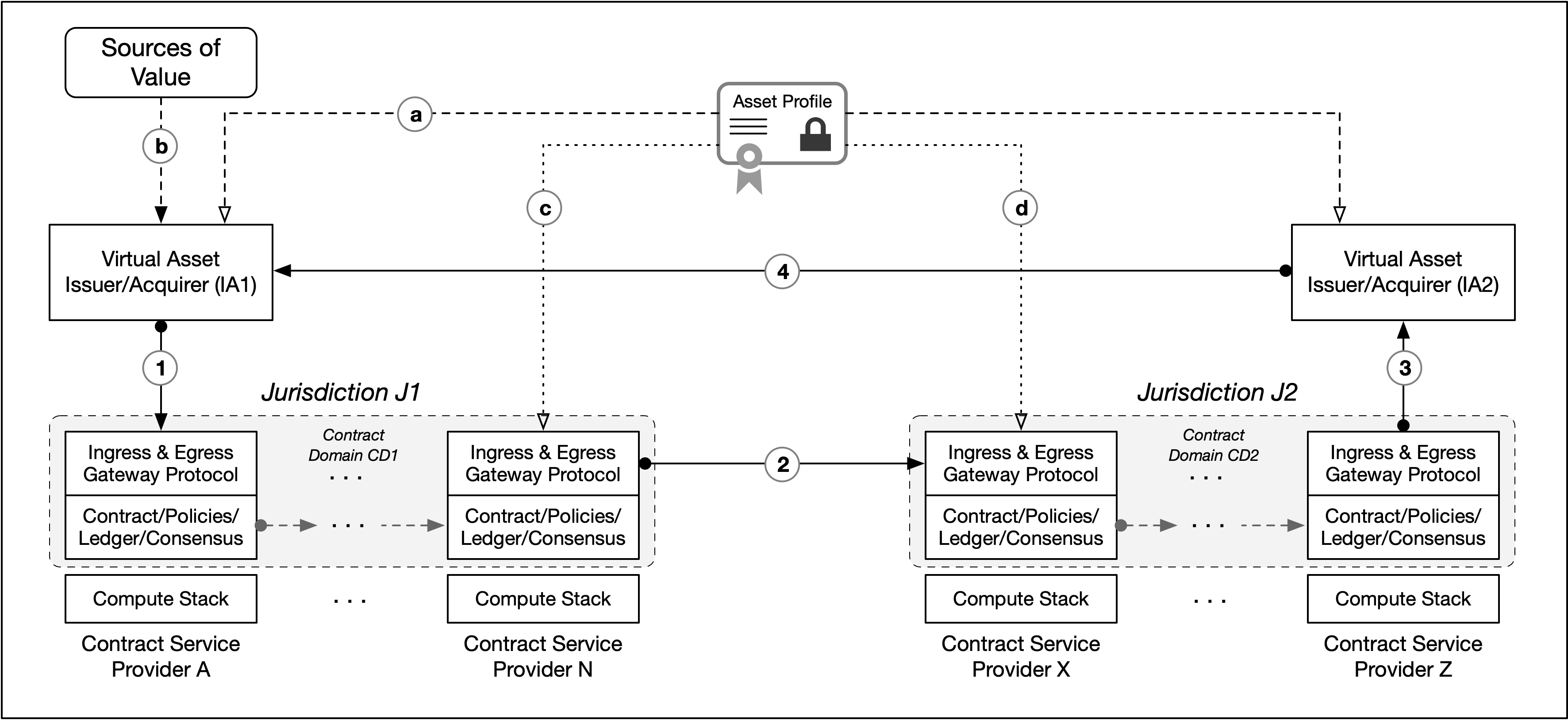}
\caption{Standard asset profiles and sources of value for virtual assets external to the contract-domain}
\label{fig:asset-profile}
\end{figure}

Figure~\ref{fig:asset-profile} provides a high level illustration.
Here the Asset Issuer (IA1) uses the asset profile definition (Step~(a))
in order to digitally represent (``convert'')
an external source of value (Step~(b))
-- such as a basket of real-world assets --
into its blockchain-based representation (Step~1).
The Asset Issuer must ensure that the contract primitives
in contract domain CD1 support this type of virtual asset,
and that the CSPs in the domain operate within 
a regulatory jurisdiction (J1) that recognizes the asset type.
Note that from a regulatory perspective
the CSPs that form (participate in) a contract domain are considered
Virtual Asset Service Providers (VASP) 
in the sense of the FATF definition~\cite{FATF-Recommendation15-2018,FATF-Guidance-2019}.
Similarly, 
the asset Issuers and Acquirers are also VASPs under the FATF definition.

When an owner (e.g. Alice) of an instance of the virtual asset
in contract domain CD1
seeks to transfer the asset to another contract domain CD2
(e.g. to a beneficiary Bob),
the owner invokes the relevant contract that
performs the asset transfer across domains (Step~(2)).
However,
the gateway nodes in domains CD1 and CD1 that handle the cross-domain transfer
must validate that:
(i) the blockchain system in domain CD2 mechanically supports the type of asset as defined in the asset-profile metadata,
and that (ii) the jurisdiction (J2) --
under which the CSPs in domain CD2 operate -- permits
incoming virtual assets of the type defined in the profile.
Thus, both gateways at providers CSP-N (in domain CD1) and CSP-X (in domain CD2)
must have access to the asset-profile metadata file (Step~(c) and Step~(d)).
The gateways must employ an asset transfer protocol
that includes (embeds) atomic commitment~\cite{DickersonHerlihy-PODC2017,Herlihy2019-CACM}.

Our use of the terms ``issuer'' and ``acquirer'' is borrowed from the classic 
{\em 4-corners model}~\cite{IBM2000-FourCorners-patent}
in the card-payments industry (i.e. credit cards),
which has been successfully deployed for over three decades now.
In the card-payments ecosystem,
the consumer (card-holder) obtains a credit card from the Issuer,
which is often a bank or financial institution where the consumer has an account.
When the consumer uses the card at a retail merchant to purchase goods
(e.g. Point of Service (POS) terminal),
the merchant forwards the transactions details to the Acquirer,
which is typically the merchant's bank or financial institution.
The Acquirer then obtains payment from the Issuer bank
(e.g. debited from the consumer's bank account).
Thus, in Figure~\ref{fig:asset-profile} the loop in closed
in Step~(4) that represents the interaction between the Acquirer IA2 
and the Issuer IA1.

The 4-corners paradigm is useful in the context of reasoning about
the design of contract domains,
for several reasons:
\begin{itemize}
\item	{\em Separation of roles and responsibilities}: The paradigm
permits the CSP role to be separated
from the role of Asset Issuer/Acquirer,
where the CSP takes-on legal obligations and 
in validating an asset prior to accepting it into the contract domain.

\item	{\em Incorporation of jurisdictional regulations}: The paradigm 
permits the notion of global jurisdictions 
to come into the picture by recognizing that
the issuers of virtual assets may reside in 
different legal jurisdictions (e.g. countries).

\item	{\em Value-conversions external to the contract domains}:
The conversion of values between value-systems and/or monetary jurisdictions
occurs at the Issuer-Acquirer level (outside the contract domain).
This is consistent with the end-to-end principle~\cite{SaltzerReed84,Clark88} discussed
previously in Section~\ref{subsec:PrinciplesInternet}.

\end{itemize}

Note that the 4-corners paradigm does not preclude an architecture
in which the asset issuers and acquirers share a common blockchain system
for the purposes of tracking conversions to/from
sources of value into virtual assets
and tracking which asset-profiles have been utilized in the process~\cite{TradecoinRSOS2018}.

\section{Conclusions}
\label{sec:Conclusions}

The success of blockchain networks hinges on their ability to make smart contract services
affordable for everyone.
This will not be achievable if transaction costs are unpredictable 
-- something arising as a side effect from a system design that 
violates the end-to-end principle by tightly coupling
the infrastructure operating costs with the tokenization of access to end-users.

Today there is a great interest in the possible use of digital currencies
at the national level in the form of Central Bank Digital Currencies (wholesale and retail)
and fiat-backed Stablecoins.
Unless the problems of blockchain interoperability and the high costs of transactions are solved,
it is unlikely that blockchains will be used for CBDCs
and that other technical solutions will need to be developed (e.g. Chaum-based payments schemes).

In this paper we have proposed the notion of a contract service provider (CSP)
as a counterpart to the well established ISP model for Internet access.
The CSP model offers an opportunity to achieve interoperability at the smart contracts layer,
and to standardize the contract primitives for specific types of virtual assets.
Similar to groups of ISPs forming peering agreements for routing IP traffic,
groups of CSPs can form communities that makes available
well-defined smart contracts authored by CSPs in the community 
for specific types of virtual assets.

The CSP model may offer a path forward for the use of decentralized nodes
as the basis for enabling CBDC distribution and usage
at a low transaction cost to end-users.
Similar to the existing interconnected autonomous systems that form the physical Internet,
interconnected CSP Communities -- as autonomous blockchain nodes
with a high degree of cross-chain interoperability --
should become the basis for making smart contracts affordable
and independent from the incumbent blockchain platforms.


\end{document}